\documentclass[10pt]{iopart}
\usepackage{iopams}
\usepackage{graphicx}
\usepackage[T1]{fontenc}
\usepackage[dvipsnames]{xcolor}
\usepackage{upgreek}
\usepackage{hyperref}

\begin{document}

\title{Exposing the trion's fine structure by controlling the carrier concentration in hBN-encapsulated MoS$_2$}

\author{Magdalena\,Grzeszczyk$^1$, Katarzyna Olkowska-Pucko$^1$, Kenji\,Watanabe$^2$, Takashi\,Taniguchi$^3$, Piotr Kossacki$^1$, Adam\,Babiński$^1$, Maciej\,R.\,Molas$^1$}

\address{$^1$ Institute of Experimental Physics, Faculty of Physics, University of Warsaw, ul.~Pasteura 5, PL-02-093 Warsaw, Poland}
\address{$^2$ Research Center for Functional Materials, National Institute for Materials Science, 1-1 Namiki, Tsukuba 305-0044, Japan}
\address{$^3$ International Center for Materials Nanoarchitectonics, National Institute for Materials Science,  1-1 Namiki, Tsukuba 305-0044, Japan}
\ead{magdalena.grzeszczyk@fuw.edu.pl}

\begin{abstract}

Atomically thin materials, like semiconducting transition metal dichalcogenides, are highly sensitive to the environment. This opens up an opportunity to externally control their properties by changing their surroundings. In this work, high-quality van der Waals heterostructures assembled from hBN-encapsulated monolayer MoS$_2$ are studied with the aid of photoluminescence, photoluminescence excitation, and reflectance contrast experiments. We demonstrate that carrier concentration in MoS$_2$ monolayers, arising from charge transfer from impurities in the substrate, can be significantly tuned within one order of magnitude by the modification of the bottom hBN flake thickness. The studied structures, characterized by spectral lines approaching the narrow homogeneously broadened limit enabled observations of subtle optical and spin-valley properties of excitonic complexes. Our results allowed us to resolve three optically-active negatively charged excitons in MoS$_2$ monolayers, which are assigned to the intravalley singlet, intervalley singlet, and intervalley triplet states.

\end{abstract}

\ioptwocol

\section{Introduction}

Two-dimensional (2D) layered materials remain for the past decade an object of great attention due to their physical properties. They are laterally bound by strong covalent bonds, which provide high in-plane stability. Whereas weak van der Waals (vdW) interlayer interactions allow isolating thin flakes up to single layer thickness.~\cite{li2016heterostructures,geim2013van} One of the most studied families of layered materials are semiconducting transition metal dichalcogenides (S-TMDs) with formula MX$_2$ where M=Mo or W and X=S, Se or Te. As the thickness of S-TMDs approaches the atomic limit, substrate-induced effects predominate their optical and electronic behavior.~\cite{guo2015charge, kang2017origin, borghardt2017engineering, ghatak2011nature, molas2019energy} In particular, it has been shown that variations in morphology, thickness, and flake-substrate bonding strength on the same substrate can still have a major impact on the material properties.~\cite{sercombe2013optical,wang2012understanding, Grzeszczyk2020}

To overcome some of those substrate-related issues, attention has been directed towards the fabrication of vdW heterostructures using hexagonal boron nitride (hBN).~\cite{Withers2015,Binder2017} The inert, flat, and uniform surface of the hBN thin flakes provides an ideal substrate for S-TMD monolayers (MLs). Their similar crystalline structure, retained by weak van der Waals interactions between individual layers, without dangling bonds on surfaces, allows for easy assembly of lattice mismatch-free structures. The hBN layers also serve as barriers preventing charge transfer from impurities $e.g.$ in commonly used SiO$_2$/Si substrates.~\cite{novoselov20162d, illarionov2016role, dolui2013origin} 
  
The MoS$_2$ MLs, like other atomically thin materials, are highly sensitive to their surroundings. It was demonstrated that employing Si or SiO$_2$ substrates led to the quenching of the photoluminescence (PL) as well as to doping effects due to the charge transfer from the substrate to the ML.~\cite{guo2015charge, buscema2014effect,  yang2018effect} The hBN flakes used as substrates proved to be remarkably suitable to prevent those effects. Their nearly charge-neutral nature with atomically flat surfaces provides buffer layers, which preserve intrinsic properties of ML MoS$_2$. The hBN substrate does not contribute to the inhomogeneous broadening of the optical transitions, often caused by surface roughness. There are also no charged impurities in hBN layers, which could generate substantial disorder in the host material.~\cite{cadiz2017excitonic, jakubczyk2019coherence} Despite the important role the hBN layers play in preserving the MoS$_2$ inherent characteristics, understanding the influence of the thickness of the bottom hBN flake on the properties of S-TMD MLs requires more in-depth exploration.
  
In this work, we conduct a systematic characterization of the optical response of the hBN-encapsulated MoS$_2$ with varying thickness of the bottom hBN layer ranging from 4~nm to 134~nm using the PL, photoluminescence excitation (PLE), and reflectance contrast (RC) experimental techniques. We show that the intensity ratio of the emission due to neutral and charged excitons in MoS$_2$ ML strongly depends on the bottom hBN flake thickness. We conclude that the carrier density in the MoS$_2$ ML can be significantly controlled by adjusting the thickness of the supporting hBN layer, particularly, in the limit of few nm. The high quality of our structures, confirmed by small linewidths of the neutral exciton emission (on the level of a few meV), allows us to unveil the fine structure of negative trions in MoS$_2$ MLs. We attribute their components to three configurations of negatively charged excitons: $i.e.$ an intravalley singlet, an intervalley singlet, and an intervalley triplet. 

\section{Results}


\subsection{Modification of carrier concentration in MoS$_2$ monolayers}

\begin{figure*}
\centering
    \includegraphics[width=\linewidth]{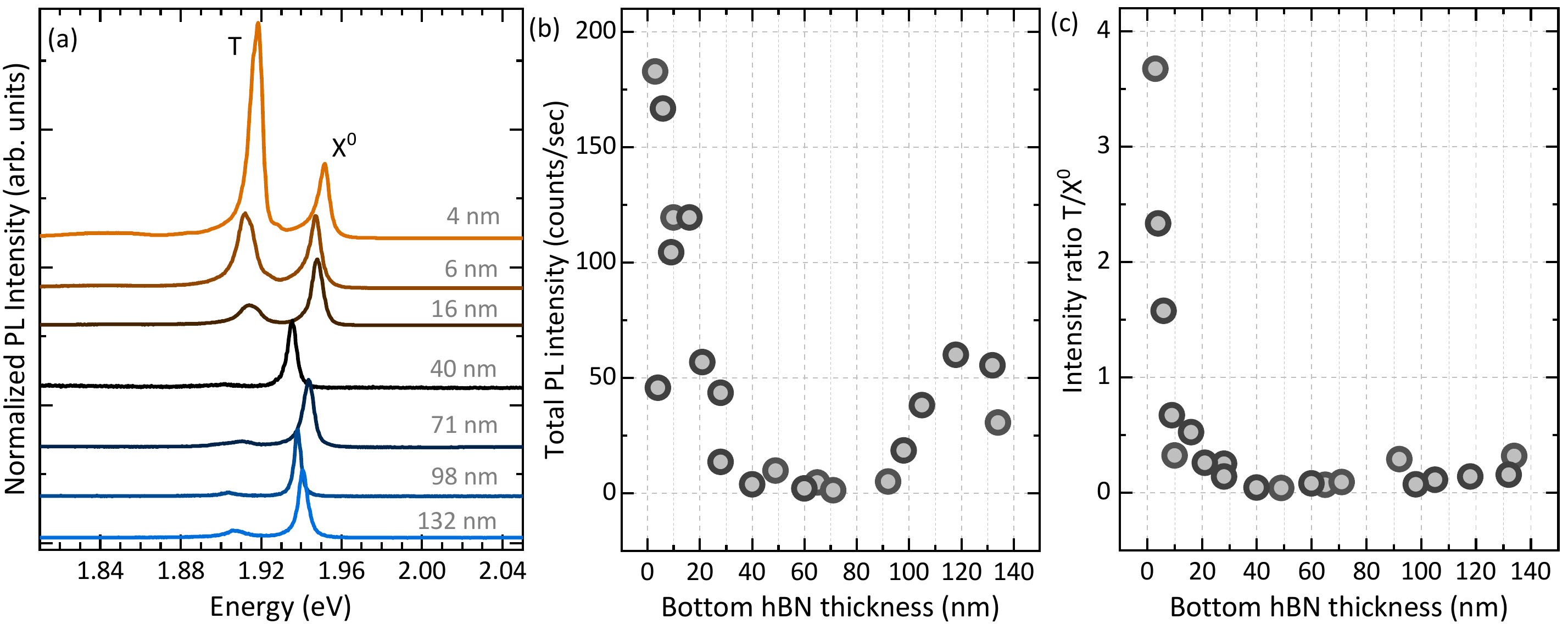}
    \caption{(a) Normalized PL spectra of hBN-encapsulated monolayer MoS$_2$ measured at $T$=5~K as a function of the bottom hBN flake thickness. The spectra are normalized to the maximum intensity of the neutral exciton emission (X$^0$) for clarity. (b) The integrated intensity of the photoluminescence (c) intensity ratio from trion and neutral exciton contributions (T/X$^0$) vs bottom hBN thickness extracted from spectra presented in (a).}
    \label{fig:PL_botthBN_thickness}
\end{figure*}

A series of samples were prepared with MoS$_2$ MLs encapsulated in hBN flakes with different thicknesses of the bottom hBN flakes ranging from 4~nm to 134~nm. Selected PL spectra of the hBN-encapsulated MoS$_2$ MLs measured at $T$=5~K are shown in Fig.~\ref{fig:PL_botthBN_thickness}(a). As can be appreciated in the Figure, two emission lines are apparent in all the presented spectra, which is consistent with several previous works done on MoS$_2$ MLs embedded between two hBN flakes.~\cite{cadiz2017excitonic, Robert2018, jadczak2020fine} Following those reports, the highest energy emission peak (X$^0$) is attributed to the neutral exciton formed in the vicinity of the optical band gap (A exciton), while the lower-in-energy feature (T) corresponds to the recombination of a negatively charged exciton (negative trion). The sign of free carriers is determined from the fine structure of the trion apparent in the PL, PLE, and RC spectra, which is analyzed in more detail in the next Section. As it is seen in Fig.~\ref{fig:PL_botthBN_thickness}(b), the total PL intensity varies significantly by more than two orders of magnitude with the thickness of bottom hBN. Moreover, the presented dependence is not monotonic, $e.g.$ the minimum of the intensities is seen for thickness of bottom hBN in the range 40 -- 90~nm. This behaviour is attributed to the interference effects in the investigated dielectric stacks in which the variation of the bottom hBN layer thickness on Si substrate can be treated as an analog of the change of SiO$_2$ thickness in commonly used SiO$_2$/Si substrates. It was demonstrated that the X$^0$ emission in MoS$_2$ MLs measured at room temperature could be enhanced about 20 times by using SiO$_2$ with thickness in the range of 192 – 328~nm. \cite{Zhang2015b} The dielectric environment is also responsible for a shift in the energy of the emission lines. A similar effect can be obtained by strain. MoS$_2$ flake is normally subjected to residual stresses resulting from the transfer procedure during sample fabrication. \cite{castellanos2013local,yan2012raman} However, in our analysis we disregard these factors, as they should have no substantial impact on doping effects, excluding charge trapping or creating localized states, as their occurrence are relatively rare.


\begin{figure*}
    \centering
    \includegraphics[width=\linewidth]{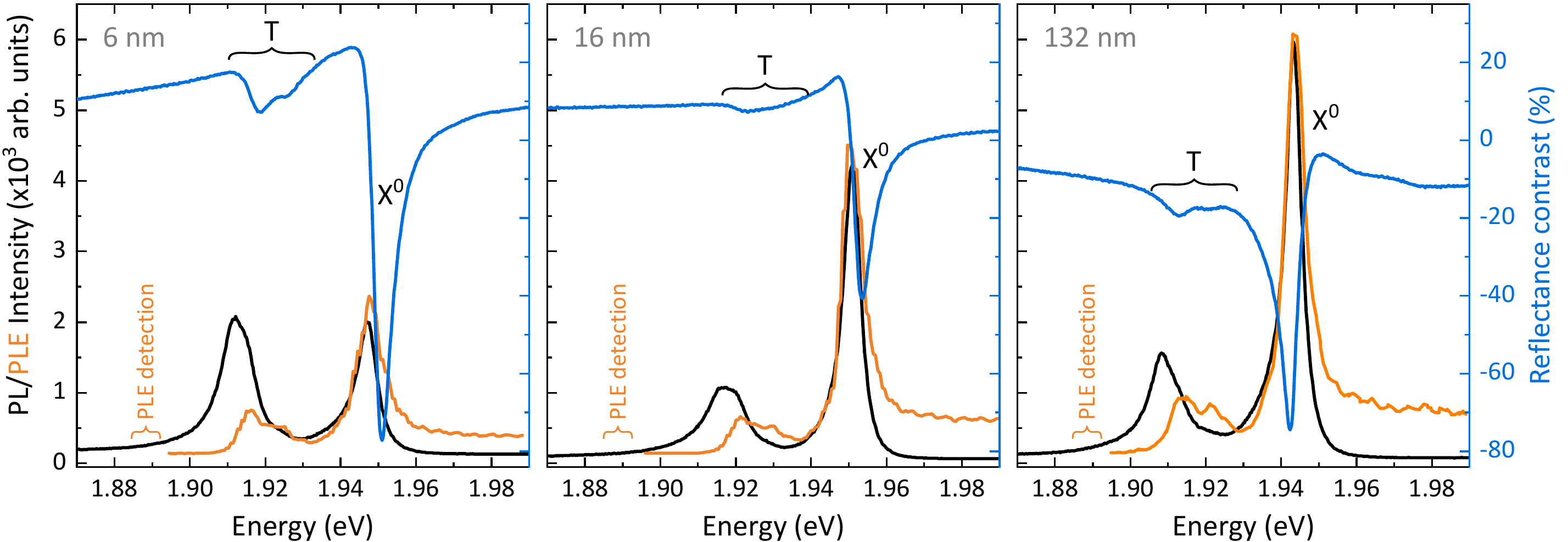}
\caption{Selected low temperature PL/PLE and RC spectra of hBN encapsulated monolayer MoS$_2$ with the thickness of the hBN substrate equal to 6~nm, 16~nm, 132~nm. }
    \label{fig:PL_selected}
\end{figure*}

Now let us focus on the detailed analysis of the measured PL spectra. For the thinnest bottom hBN flake of 4~nm thickness, the T emission is much more intense as compared to the X$^0$ one. When the bottom hBN flake gets thicker, the optical emission due to the neutral exciton starts to dominate the PL spectra, while the trion intensity substantially decreases. Fig.~\ref{fig:PL_botthBN_thickness}(c) demonstrates the intensity ratio of the X$^0$ and T emissions as a function of hBN thickness. One can observe a decline in this ratio by two orders of magnitude from $\sim$3 to $\sim$0.03 with the hBN thickness approaching approx. 20~nm. For thicker hBN flakes the intensity ratio saturates to a roughly constant value. This behavior suggests the influence of external factors on the emission spectra and thus on the inherent properties of the studied MLs.

MoS$_2$ crystals generally bear inherent $n$-type doping, including the MLs discussed within this report. This characteristic is linked to abundant sulfur vacancies in the bulk crystal. \cite{lin2016defect,saha2018native,hong2015exploring, yu2014towards,liu2013sulfur} On the other hand, many reports show that the properties of this material vary depending on the structure in which it is used. In particular, encapsulating MoS$_2$ significantly changes the optical or electrical performance of MLs, indicating their sensitivity to some external factors that affect their intrinsic properties. \cite{baugher2013intrinsic,wang2012integrated,radisavljevic2013mobility} 

In our analysis, we adopt the reasoning that MLs obtained from a single bulk crystal share similar properties, particularly, a comparable level of doping. This means that the observed variation in the PL response within a given series of samples can be associated with the variation of the bottom hBN flake thickness. The distinctly higher emission intensity of the T line compared to the X$^0$ one for the thinnest lower hBN flake is a signature of the high electron concentration in these MoS$_2$ MLs. We belive that the increased concentration stems from quantum tunneling of carriers from impurities present in SiO$_2$/Si substrates. 

As similar considerations have not been widely addressed in optical measurements, we invoke the results of electrical studies. Since hBN is an insulator with a large energy band gap, it is an excellent dielectric material that has been widely used as a building block of high-performance devices, $i.a.$ as a tunneling barrier. \cite{young2012electronic,britnell2012electron,britnell2012field,iqbal2018gate,mishchenko2014twist}
Even though some reports suggest that a single atomic hBN layer acts as effective isolation, \cite{britnell2012electron} it is often found, that the very thin flakes do not provide efficient barriers for carriers. \cite{illarionov2016role,dolui2013origin}. As the transmission probability of the hBN barrier decreases exponentially with the number of atomic layers, most devices rely on the barrier of single nanometers in thickness. \cite{iqbal2018gate, jung2017direct, paul2020modulating,iqbal2019fowler,chu2017selective} However, the electron wave function can penetrate even larger number of layers, up to 9 nm thick. \cite{britnell2012field,kang2016effects,lee2011electron,kim2018spin} 

A parallel pattern can be followed in our results, where charge carrier tunneling seems to occur for structures with bottom hBN thickness below $\sim$10 nm, as manifested by the strong trion-related emission (see Fig.\ref{fig:PL_botthBN_thickness}). Above this limit, the density of free carriers in the MoS$_2$ MLs attains an inherent doping level of a natural crystal. 

To provide an estimate of the carrier concentration in the investigated MLs, the extracted values of X$^0$--T ratio, presented in Fig.~\ref{fig:PL_botthBN_thickness}(c), are compared with the PL results measured as a function of free electron concentration in Ref.~\cite{klein2020quantized}. It has been found that the electron concentration ($n_e$) equals about 3 $\times$ 10$^{12}$ cm$^{-2}$ in the MLs on 4~nm hBN, then decreases by about one order of magnitude for $\geq$ 10 nm thick bottom hBN flake. The presented approach can be adapted to other S-TMDs MLs, such as MoSe$_2$, MoTe$_2$, WS$_2$, and WSe$_2$, which gives us a useful tool to tune the carrier concentration in S-TMD MLs deposited on hBN flakes.


\subsection{Fine structure of negative trions}\label{sec:trion}

To investigate thoroughly the effect of bottom hBN thickness on both the emission and absorption processes in MoS$_2$ MLs,  we have focused our further investigations using three experimental techniques (PL, PLE, and RC) on selected samples with hBN substrate thicknesses of 6~nm, 16~nm, and 132~nm, see Fig. \ref{fig:PL_selected}. As can be appreciated in the Figure, the presence of the X$^0$ and T resonances are observed in all of the measured spectra. Let us turn the attention to a detailed analysis of the feature in the vicinity of the T complex. While two absorption-type resonances can be easily distinguished for the negative trions in the PLE and RC spectra, the corresponding emissions comprise up to three components. A double structure of the trion in MoS$_2$ ML has been recently ascribed to the intravalley spin-singlet and intervalley spin-singlet negatively charged excitons for the higher and lower-energy features, respectively.~\cite{jadczak2020fine, Roch2019} The lowest energy resonance visible only in the PL spectra has been related to intervalley spin-triplet negatively charged exciton (see the next Section for details). 


It is also worth pointing out that the interference effects in the investigated dielectric stacks affect not only the overall PL intensity (see Fig.~\ref{fig:PL_botthBN_thickness}), but also the lineshape of resonances observed in the RC spectra (see Fig.~\ref{fig:PL_selected}). A similar effect on the observed RC features was reported for MoS$_2$ MLs deposited on SiO$_2$/Si substrates. \cite{Li2018}

\begin{figure*}
    \centering
    \includegraphics[width=\linewidth]{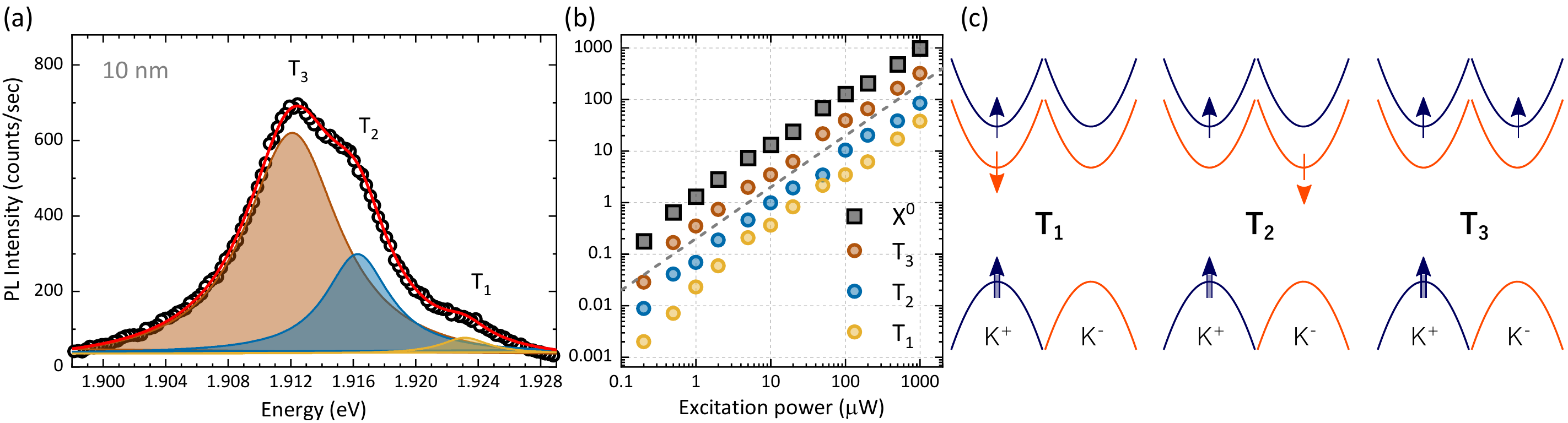}
\caption{(a)  Photoluminescence spectrum of negative trions measured on the MoS$_2$ monolayer with 10~nm bottom hBN. The coloured Lorentzians display fits to the corresponding T$_1$, T$_2$, and T$_3$ lines. (b) Integrated intensities of the negative trions and the neutral exciton as a function of excitation power. The dashed grey line is a guide to the eye indicating a linear increase. (c) Schematic illustration of possible spin configurations for optically-active (bright) negatively charged excitons formed in the K$^+$ point in the monolayer MoS$_2$. The T$_1$, T$_2$, and T$_3$ states correspond to intravalley spin singlet, intervalley spin singlet, and intervalley spin triplet. Complexes for which a hole is located at the K$^+$ point of the Brillouin zone are depicted. }
    \label{fig:trions}
\end{figure*}

The fine structure of the negative trion, apparent in both emission and the absorption experiments, is displayed distinctly in Fig.\ref{fig:trions}(a), in which a combination of three Lorentzian curves was fitted to data. The extracted energy separation between the T$_1$ and T$_2$ is equal to $\sim$7~meV, which is consistent with the corresponding value of around 8~meV from RC measurements.~\cite{Roch2019} The extracted T$_2$-T$_3$ energy separation is of the order of 4~meV. The observation of the T$_3$ line in the PL spectrum is most surprising, as its contribution to the RC and PLE spectra is absent (see Fig.~\ref{fig:PL_selected}. To verify the origin of the T lines, the  PL spectra have been measured with varying excitation power. The integrated intensities of the negative trions and the neutral exciton display a linear dependence on the excitation power, see Fig.~\ref{fig:trions}(b), as is expected for excitonic features comprising a single electron-hole ($e$-$h$) pair.~\cite{Klingshirn2012}

The apparent fine structure of the negative trion might seem surprising, as the theoretically predicted structure of the conduction (CB) and valence (VB) bands leads to the optical activity of the energetically lowest transition.~\cite{Kormanyos2015} This picture however does not include the excitonic character of the optical response of S-TMD MLs.~\cite{molas2019energy} It has recently been demonstrated that the excitonic ground state in MoS$_2$ MLs encapsulated in hBN is optically inactive or dark,~\cite{Robert2020} which places this material within the family of so-called $darkish$ MLs. For other darkish MLs, such as WS$_2$ and WSe$_2$, two configurations of the negative trions, $i.e.$ intravalley spin-singlet and intervalley spin-singlet states, were reported in the literature.~\cite{courtade2017charged, Vaclavkova2018, Lyons2019, Kapuscinski2020} It is important to mention that the small CB spin-orbit splitting in MoS$_2$ MLs, predicted to equal about 3~meV,~\cite{Kormanyos2015} leads to yet another effect. This is the reordering of the spin-split and spin-polarized CB subbands in the K valley occupied by an electron-hole ($e$-$h$) pair forming a trion. The reordering results from a difference in the effective electron mass in the subbands combined with the $e$-$h$ interaction (see Ref.~\cite{jadczak2020fine, Roch2019} for details). This leads to the same arrangement of CB subbands in both the K$^+$ and K$^-$ points of the Brillouin zone, in contrast to the opposite arrangement in the W-based MLs.~\cite{courtade2017charged, Vaclavkova2018, Lyons2019, Kapuscinski2020}. Consequently, three spin configurations of the optically-active (bright) negatively charged excitons in MoS$_2$ MLs may be formed in the K$^+$ valley, $i.e.$ intravalley spin singlet (T$_1$), intervalley spin singlet (T$_2$), and intervalley spin-triplet (T$_3$), see Ref.~\cite{druppel2017diversity}. Fig.~\ref{fig:trions}(c) presents schematic illustration of the discussed possible spin configurations for optically-active (bright) negative trions formed in the K$^+$ point in the monolayer MoS$_2$. Note that all corresponding negative trions can also be formed in the K$^-$ valley, which leads to two possible conﬁgurations of a given complex. 

By comparing the measured RC, PLE, and PL spectra on the MoS$_2$ MLs with 6~nm and 16~nm bottom hBN flakes (see Fig.~\ref{fig:PL_selected}) and the RC spectra reported in Ref.~\cite{Roch2019}, we relate the highest- (T$_1$) and middle-energy (T$_2$) resonances to the intravalley spin-singlet and intervalley spin-singlet negative trions, respectively. The assignment of the T$_3$ is the most difficult, as its contribution is only manifested in PL spectra and it has not been reported so far. We ascribe tentatively T$_3$ line to the emission of the intervalley spin-triplet state of the charged exciton, see Fig.~\ref{fig:trions}(c). The intervalley trion is the energetically lowest, as the two electrons occupy both valleys allowing them to reside in the energetically lowest CB.~\cite{druppel2017diversity} The appearance of the T$_3$ emission can be due to a photo-doping process in the PL experiment, which can explain the lack of the T$_3$ contribution in the corresponding RC and PLE spectra. Recently, it has been demonstrated that the laser excitation may induce free electron concentration in MoS$_2$ MLs ranging up to several times 10$^{12}$ cm$^{-2}$.~\cite{Gadelha2020} Due to the small CB spin-orbit splitting and the photo-doping processes, we assumed that the Fermi level is high enough to provide free electrons on the higher energy CB subband allowing the formation of the intervalley triplet state of the negative trion.



\begin{figure}
    \centering
    \includegraphics[width=\linewidth]{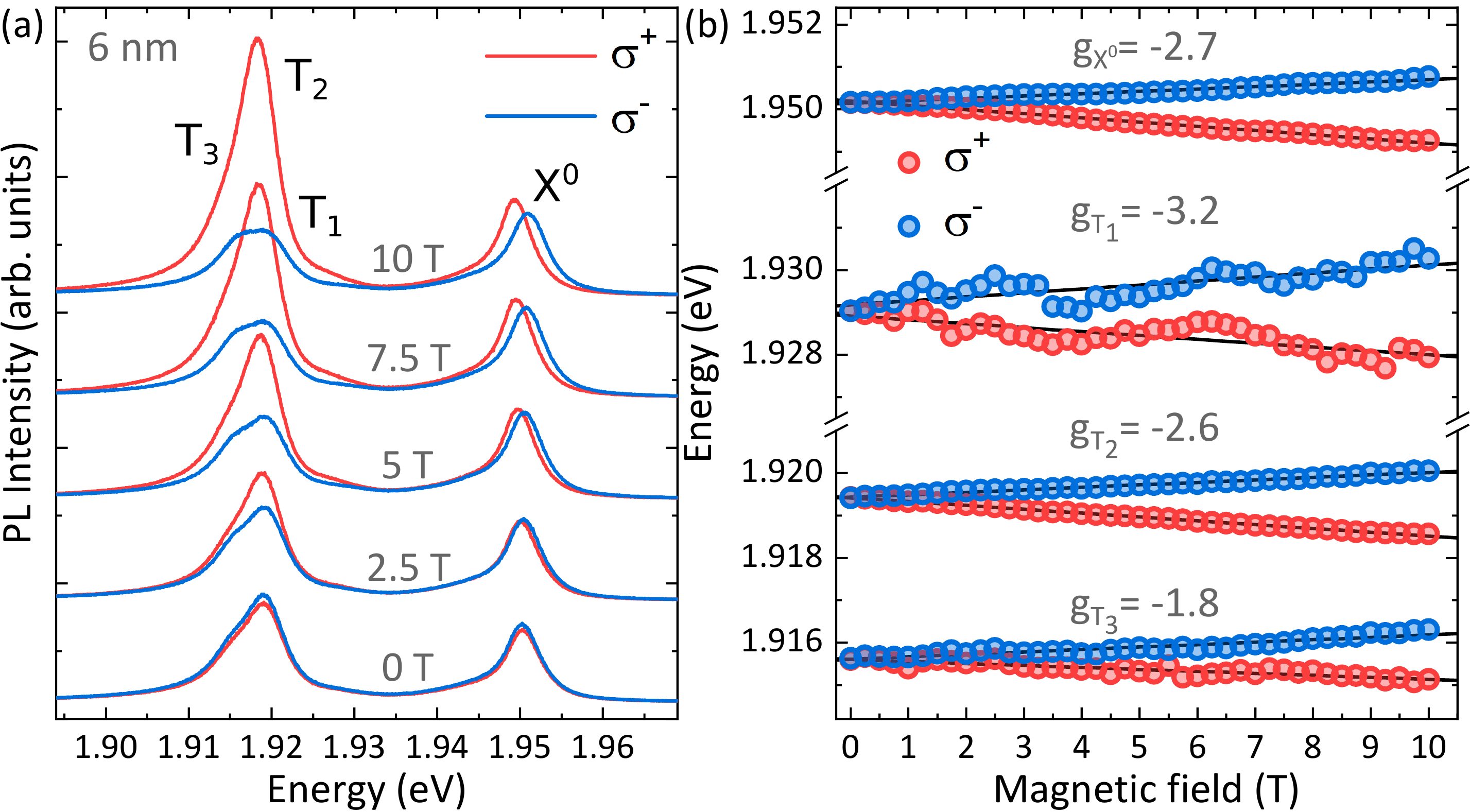}
\caption{(a) Helicity-resolved PL spectra of MoS$_2$ monolayer with 6~nm bottom hBN at selected values of the applied out-of-plane magnetic field. The red (blue) colour corresponds to the $\sigma^+$ ($\sigma^-$) polarized spectra. The spectra are vertically shifted for clarity. (b) Transition energies of the excitonic complexes extracted from the fits to the PL spectra as a function of magnetic field. The solid black curves represent fits with Eq.~\ref{equ}.}
    \label{fig:trionsB}
\end{figure}

In order to investigate the properties of the negative trions, the helicity-resolved PL spectra were measured on the MoS$_2$ ML on the 6~nm bottom hBN flake in magnetic fields up to 10~T oriented perpendicularly to ML's plane (out-of-plane configuration), see Fig.~\ref{fig:trionsB}(a). Applying an out-of-plane magnetic field yields the excitonic Zeeman effect,~\cite{Koperski_2018} which is apparent as splitting into two circularly polarized components of each transition. To facilitate the analysis of that evolution, we fitted these results with Lorentzian functions. The extracted transitions energies of both the $\sigma^+$ and $\sigma^-$ polarized components as a function of magnetic fields are summarized in Fig.\ref{fig:trionsB}(b). As can be appreciated in the Figure, all the $\sigma^+$/$\sigma^-$ features experience redshift/blueshift with increasing magnetic field. The energy evolution of the $\sigma^\pm$-polarized lines, $E_{\sigma^{\pm}}$ can be described by the following equation:
\begin{equation}
\hspace{20mm}E_{\sigma^{\pm}}(B)=E_0\pm \frac{1}{2} g\mu_BB,
\label{equ}
\end{equation}
\noindent where $E_0$ is the transition energy at zero magnetic field, $g$ denotes the $g$-factor of the considered excitonic complex and $\mu_B$ is the Bohr magneton.  Results of the fitting are shown in Fig.\ref{fig:trionsB}(b) with solid black curves. The obtained $g$-factor of the neutral exciton (X$^0$) equals about -2.7. This value is consistent with previously reported $g$-factor values for the X$^0$ line measured in MoS$_2$ MLs encapsulated in hBN, which stay within the range from -1.9 to -3.~\cite{cadiz2017excitonic, jadczak2020fine, klein2020quantized, goryca2019revealing} For the trion-related transitions: T$_1$, T$_2$, and T$_3$, the corresponding $g$-factors are equal to -3.2, -2.6, and -1.8. In spite of the clear decrease of the trion $g$-factor with decreasing the transition energy, the found difference cannot be explained with the existing theories for excitonic $g$-factor in S-TMD MLs and requires further theoretical investigation.

Apart from the aforementioned analysis of the magnetic-field dependence of transition energies, the applied magnetic field may also affect the emission intensities of the $\sigma^\pm$ components.~\cite{Kapuscinski2020, Koperski_2018} As can be seen in Fig.~\ref{fig:trionsB}(a), both $\sigma^\pm$ components of the X$^0$ line are characterized by almost the same intensity in magnetic fields, which is characteristic for this complex in darkish MLs.~\cite{Koperski_2018,molas2019energy,Roch2019} Simultaneously, all trion-related emissions are polarized in higher fields. This effect is seen distinctly for the T$_1$ and T$_2$ line, for which the lower/higher energy ($\sigma^+$/$\sigma^-$) component gains/loses in intensity with the increased magnetic field. The observed evolution for T$_1$ and T$_2$ line is analogous for observed in RC spectra (see Supplementary Materials) and was previously reported in Ref.~\cite{Roch2019}. This effect can be understood in terms of thermal distribution of free electrons between field-split states (see Refs.~\cite{Roch2019, Koperski_2018} for details). In the case of the T$_3$ line, the opposite behaviour is expected so that the magnetic field induces the gain/loss of higher/lower energy ($\sigma^-$/$\sigma^+$) component.~\cite{druppel2017diversity} Unfortunately, the quality of results, presented in Fig.~\ref{fig:trionsB}(a), does not allow us to unquestionably assign it. Moreover, the  understanding of the magnetic-field evolution of the T$_3$ intensity can be more complex, as the emission of this complex is observed due to the photo-doping effect. It should be also emphasized that the magnetic field dependence of trion intensities in MoS$_2$ ML is completely different from the corresponding one observed for intravalley and intervalley negatively charged excitons in WS$_2$ ML.~\cite{Kapuscinski2020}

\section{Summary}

In summary, we present a systematic study of the effect of lower hBN thickness on the optical properties of hBN encapsulated MoS$_2$ MLs. Our results demonstrate that high electron concentration in TMD MLs usually originates from charge transfer from the substrate. The hBN flakes used as substrates act as barriers for carriers originating from those impurities. The effectiveness of such a barrier depends on its thickness. This approach can be used to manipulate the carrier concentration in the studied material. The reported investigation also allows unraveling the fine structure of the trion-related transition in the MoS$_2$ MLs. The three optically active complexes have been attributed to an intravalley singlet, an intervalley singlet, and an intervalley triplet. The observation of the latter transition in the emission is related to the small CB spin-orbit splitting in MoS$_2$ ML and the photo-doping processes. The effective g-factors of all trion complexes have been determined. The existence of these trion features confirms the reordering of the spin-split and spin-polarized CB subbands in the K valley, allowing the three spin configurations of the bright negatively charged excitons. Given that substrates are typically needed for most research and applications, our results are highly relevant allowing for a deeper qualitative understanding of environmentally induced changes on intrinsic properties of S-TMDs MLs.

Sample fabrication, experimental details and reflectance contrast results in magnetic field can be found in Supplementary Material.

\section*{Acknowledgements}
The work has been supported by the National Science Centre, Poland (grant no. 2017/27/B/ST3/00205, 2017/27/N/ST3/01612, 2018/31/B/ST3/02111), EU Graphene Flagship project (no.785219), the ATOMOPTO project (TEAM programme of the Foundation for Polish Science, co-financed by the EU within the ERD-Fund), and the CNRS via IRP "2DM" project. K. W. and T. T. acknowledge support from the Elemental Strategy Initiative conducted by the MEXT, Japan, (grant no. JPMXP0112101001), JSPS KAKENHI (grant no. JP20H00354), and the CREST (JPMJCR15F3), JST.

\qquad

\bibliographystyle{unsrt}
\bibliography{bib}

\end{document}


\title{Exposing the trion's fine structure by controlling the carrier concentration in hBN-encapsulated MoS$_2$ \\ Supplementary Material}

\email{Magdalena.Grzeszczyk@fuw.edu.pl}

\author{Magdalena\,Grzeszczyk} 
\author{Katarzyna Olkowska-Pucko}
\affiliation{Institute of Experimental Physics, Faculty of Physics, University of Warsaw, ul.~Pasteura 5, PL-02-093 Warsaw, Poland}
\author{Kenji\,Watanabe}
\affiliation{Research Center for Functional Materials, National Institute for Materials Science, 1-1 Namiki, Tsukuba 305-0044, Japan}
\author{Takashi\,Taniguchi}
\affiliation{International Center for Materials Nanoarchitectonics, National Institute for Materials Science,  1-1 Namiki, Tsukuba 305-0044, Japan}
\author{Piotr Kossacki}
\author{Adam\,Babiński}
\author{Maciej\,R.\,Molas}
\affiliation{Institute of Experimental Physics, Faculty of Physics, University of Warsaw, ul.~Pasteura 5, PL-02-093 Warsaw, Poland}

\begin{abstract}
\end{abstract}

\maketitle

\section{SAMPLE FABRICATION}

The investigated MoS$_2$ monolayers and hBN flakes were fabricated by two-stage PDMS-based mechanical exfoliation of naturally occurring bulk crystal. Initially, the hBN thin films were exfoliated onto a 90 nm SiO$_2$/Si substrate and annealed at 200$^{\circ}$C. That non-deterministic approach provides the best quality of the substrate surface. Subsequent layers were transferred deterministically using a microscopic system equipped with a $x$-$y$-$z$ motorized positioners. The complete structures were annealed at 160$^{\circ}$C for 1.5 hour in order to ensure the best layer-to-layer and layer-to-substrate adhesion and to eliminate a substantial portion of air pockets on the interfaces between the constituent layers.

\section{EXPERIMENTAL SETUP}

The PL, PLE, and RC experiments were performed using a \mbox{$\lambda$=515~nm} (2.41~eV) continuous wave (CW) laser diode, a tunable CW single-mode DCM dye laser, and a 100 W tungsten halogen lamp, respectively. The studied samples were placed on a cold finger in a continuous flow cryostat mounted on $x$-$y$ manual positioners. The excitation light was focused by means of a 100x long-working distance objective with a 0.55 numerical aperture producing a spot of about 1 $\upmu$m diameter in PL/PLE measurements and 4 $\upmu$m diameter in RC measurements. The signal was collected via the same microscope objective, sent through a 0.75 m monochromator, and then detected by using a liquid nitrogen cooled charge-coupled device camera. The excitation power focused on the sample was kept at 70~$\upmu$W during all measurements to avoid local heating, except for experiments as a function of excitation power.

Magneto-optical experiments were performed in the Faraday configuration using a free-beam-optics arrangement in a superconducting coil producing magnetic fields up to 10 T.  Samples were placed in a helium gas atmosphere at the temperature of 5 K. The excitation light was focused by a single aspheric lens to a spot smaller than 2 $\upmu$m in case of laser and 6 $\upmu$m for the white light. The lens was mounted on the x-y-z piezoelectric stage inside the cryostat. The circular polarizations of the emission were analyzed using a polarizer and a $\lambda$/4 waveplate placed directly in front of the spectrometer.

\section{Refelctance contrast in magnetic field}
\begin{figure}[b]
\centering
\includegraphics[width=.9\linewidth]{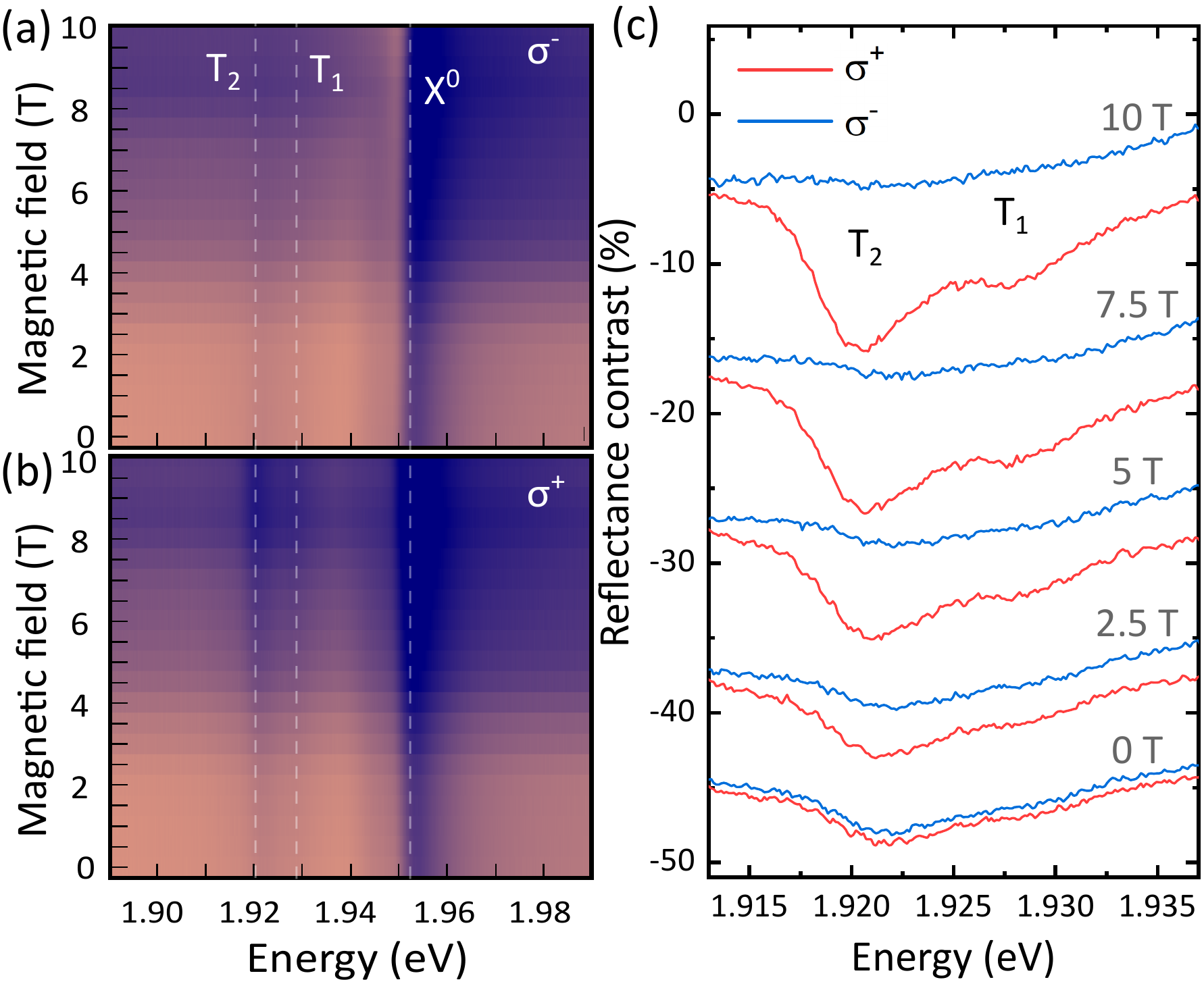}
\caption{False-color map of the helicity-resolved low temperature reflectance contrast measured on the encapsulated MoS$_2$ monolayer with 6~nm bottom hBN for (a)
$\sigma^-$ (b) $\sigma^+$ polarized light. The white dashed lines are indicated as a guide to the eye. (c) Reflectance contrast spectra of trion-related emission at selected magnetic fields. The red (blue) colour correspond to the $\sigma^+$ ($\sigma^-$) polarized light. The spectra are vertically shifted for clarity.}\label{fig:RC}
\end{figure}

Due to the complexity of resolving the individual trion-releated emission features in helicity-resolved PL spectra shown in main text, we additionally performed reflectance measurements. As mentioned in the main text: two absorption-type resonances can be easily distinguished for the negative trions in the RC spectra, T$_1$ and T$_2$, ascribed to the intravalley spin-singlet and intervalley spin-singlet, respectively.~\cite{druppel2017diversity, jadczak2020fine, Roch2019}

As presenteed in Fig.~\ref{fig:RC}, similar to results of PL experiments, with $\sigma^-$ polarization as the magnetic field increases, the features associated with trions weaken to the point of being indistinguishable. In the opposite photon polarization, $\sigma^+$ both resonances gain intensity with increasing magnetic field. The observed evolution for T$_1$ and T$_2$ line is analogous for observed in RC spectra reported in Ref.~\cite{Roch2019}.

\bibliography{bib}
\bibliographystyle{apsrev4-1}